\documentclass[a4paper,twocolumn,showpacs,showkeys,english]{revtex4}
\usepackage{amsmath} 
\usepackage{amssymb} 
\usepackage[dvipdfm]{graphicx} 

\newcommand {\DS} {\displaystyle} 

\begin{document}

\title{Finite amplitude method for the quasi-particle-random-phase approximation}

\author{Paolo Avogadro}
\affiliation{RIKEN Nishina Center, Wako-shi, 351-0198, Japan}
\affiliation{
Dipartimento di Fisica, Universit\`a degli Studi di Milano, via Celoria 16, 20133 Milan, Italy}
\affiliation{
INFN, Sezione di Milano, Milan, Italy}
\author{Takashi Nakatsukasa}
\affiliation{RIKEN Nishina Center, Wako-shi, 351-0198, Japan}
\affiliation{
Center for Computational Sciences, University of Tsukuba,
Tsukuba, 305-8571, Japan}

\date{\today}

\begin{abstract}
We present the finite amplitude method (FAM), originally
proposed in Ref. \cite{fam}, for superfluid systems.
A Hartree-Fock-Bogoliubov code may be transformed into a code
of the quasi-particle-random-phase approximation (QRPA)
with simple modifications.
This technique has advantages
over the conventional QRPA calculations,
such as coding feasibility and computational cost.
We perform the fully self-consistent linear-response calculation
for a spherical neutron-rich nucleus $^{174}$Sn, modifying the HFBRAD code,
to demonstrate the accuracy, feasibility,
and usefulness of the FAM.
\end{abstract}


\maketitle


\begin{section}{Introduction}\label{intro}
Elementary modes of excitation in nuclei provide valuable information
about the nuclear structure. 
The random-phase approximation (RPA) based on energy density functionals (EDF)
is a leading theory applicable both to low-lying excited states and
giant resonances \cite{RS,Bender}.
Although the fully self-consistent treatment of
the residual (induced) interactions
for the realistic energy functionals is becoming more and more prevalent
\cite{Imagawa,Paar,Nakatsukasa,Terasaki1,Fracasso,Sil,Terasaki2,Artega,Peru,
Losa,Terasaki-def},
the RPA calculations for deformed nuclei 
are still computationally demanding.
At present, the quasi-particle random-phase approximation (QRPA)
for deformed superfluid nuclei are
limited only to axially deformed cases
\cite{Artega,Peru,Yoshi1,Losa,Terasaki-def,Yoshi2},
except for Ref. \cite{Ebata} with an approximate treatment of the pairing
interaction.

Recently, there has been a renewed interest in the solution of the RPA
problem \cite{fam,Inakura,DobaArnoldi}.
In Ref. \cite{fam}, the finite amplitude method (FAM) was proposed as a
feasible method for a solution of the RPA equation.
The FAM allows to calculate all the induced fields as a finite
difference, employing a computational program of the static mean-field
Hamiltonian.
Recently, the FAM has been applied to the electric dipole excitations
in nuclei using the Skyrme energy functionals \cite{Inakura}.
There has been also a calculation making use the iterative Arnoldi algorithm for
a solution of the RPA equation \cite{DobaArnoldi}.
These newly developed technologies in conjunction with fast solving
algorithms for linear systems open the possibility to explore systematically
the nuclear excitations over the entire nuclear chart.

So far, these new techniques \cite{fam,Inakura,DobaArnoldi}
have been developed for solutions
of the RPA without the pairing correlations.
It is well known, however, that almost all but magic nuclei display superfluid
features.
Therefore, a further improvement is highly desirable to make these
methods applicable to
the QRPA equations including correlations in
the particle-particle and hole-hole channels. 
The purpose of the present paper is to generalize the FAM to superfluid
systems, which enables us to perform a QRPA calculation utilizing a static
Hartree-Fock-Bogoliubov (HFB) code with minor modifications.
Our final goal would be the construction of a fast computer program for
the fully self-consistent and triaxially deformed QRPA.
This paper is a first step toward the goal, to present the basic equations
of the FAM for the QRPA and show the first result for spherical nuclei.
We use the spherically symmetric HFB code called HFBRAD
\cite{BennaDoba} to be converted into the QRPA code.

This paper is organized as follows:
In Sec. \ref{secTDHFB}, the QRPA equation is derived as the
small-amplitude limit of the time-dependent HFB (TDHFB) equations.
In Sec. \ref{secFAM}, we obtain the FAM formulae for the calculation of
the induced fields.
In Sec. \ref{sec:fammethod}, we summarize all the relevant formulae for
practical application of the FAM.
In Sec. \ref{sectComparison}, we apply the FAM to the HFBRAD and compare
the result with that of another self-consistent calculation.
Sec. \ref{conclusioni} is devoted to the conclusions.
\end{section}

\section{Small amplitude limit of the TDHFB}\label{secTDHFB}
In this section, we recapitulate the basic formulation of the TDHFB and
its small-amplitude limit.
In general, we will follow the notation in Ref.~\cite{RS}
unless otherwise specified.
We also use $\hbar=1$ in the following equations. 

We start from the energy functional $\mathcal{E}[\rho,\kappa,\kappa^*]$ which is
a functional of the density matrix and pairing tensor.
\begin{equation}
 \begin{split}
     \rho_{kl} = \langle \Phi| c^{\dagger}_l c_k | \Phi \rangle,~~~
   \kappa_{kl} = \langle \Phi| c^{       }_l c_k | \Phi \rangle ,
 \end{split}  
\end{equation}
where $| \Phi\rangle $ is the HFB state.  
The single-particle Hamiltonian $h$ and the pairing potential $\Delta$
are obtained with a variation of the energy functional
with respect to $\rho$ and $\kappa^*$, respectively.
\begin{equation}
\begin{split}
 h_{kl}[\rho,\kappa,\kappa^*]
=\frac{\partial \mathcal {E}}{\partial \rho_{lk}},
~~~
 \Delta_{kl}[\rho,\kappa,\kappa^*]
=\frac{\partial \mathcal{E}}{\partial \kappa^{*}_{kl}}.
\end{split}
\end{equation}
The Bogoliubov quasi-particles, $(a_\mu, a_\mu^\dagger)$,
have a linear connection
to the bare particles, $(c_k,c_k^\dagger)$;
$a_\mu^\dagger = \sum_{k} (U_{k\mu} c_k^\dagger + V_{k\mu} c_k )$.
Here, the index $k$ indicates
the adopted basis such as the harmonic oscillator states or
the coordinate space.
The quasi-particles $a_\mu$ are chosen
so as to diagonalize the HFB Hamiltonian \cite{RS}.
\begin{equation}
\label{HFB_Hamiltonian}
H_0 =\frac{1}{2}
\begin{pmatrix}
c^{\dagger} & c
\end{pmatrix}
\begin{pmatrix}
 h-\lambda     &    \Delta \\
 -\Delta^* & -(h^{*}-\lambda)
\end{pmatrix}
\begin{pmatrix}
 c \\
 c^{\dagger}
\end{pmatrix}
= \sum_\mu E_\mu a_\mu^\dagger a_\mu .
\end{equation}
Here, the normal ordering is assumed.

In a similar manner,
the time-dependent quasi-particles $a_\mu^\dagger(t)$
are characterized by the time-dependent wave functions $(U(t),V(t))$ by
$a_\mu^\dagger(t) = \sum_{k} \{U_{k\mu}(t) c_k^\dagger + V_{k\mu}(t) c_k \}$.
The time evolution of the quasi-particles under a one-body external
perturbation $F(t)$ are determined by
the following TDHFB equation.
\begin{equation}\label{motodist}
i \frac{\partial a_{\mu}(t)} {\partial t}=[H(t)+F(t),a_{\mu}(t)] ,
\end{equation}
where the TDHFB Hamiltonian is given by
\begin{eqnarray}
H(t) 
&=&\sum_{kl} 
\left\{ h_{kl}(t) - \lambda\delta_{kl} \right\} c_k^\dagger c_l
\nonumber \\
&& + \sum_{k>l} \left\{ \Delta_{kl}(t) c_k^\dagger c_l^\dagger
                       +\Delta_{kl}^*(t) c_l c_k \right\}
\nonumber \\
\label{TDHFB_Hamiltonian}
&=&\frac{1}{2}
\begin{pmatrix}
 c^{\dagger} & c
\end{pmatrix}
\begin{pmatrix}
 h(t)-\lambda     &    \Delta(t) \\
 \Delta^{\dagger}(t) & -(h^{*}(t)-\lambda)
\end{pmatrix}
\begin{pmatrix}
 c \\
 c^{\dagger}
\end{pmatrix}
\end{eqnarray}
Here and hereafter, the constant shift is neglected, since it
does not play any role in the TDHFB equation (\ref{motodist}).
$h(t)$ and $\Delta(t)$ become time-dependent,
since they depend on the densities,
$\rho(t)=V^*(t) V^T(t)$ and
$\kappa(t)=V^*(t) U^T(t)=-U(t)V^\dagger(t)$, which are time-dependent.
Note that the static quasi-particles correspond to a quasi-static solution of
Eq. (\ref{motodist}), $a_\mu(t)=a_\mu e^{iE_\mu t}$, with $F(t)=0$.

Let us assume that the nucleus is
under a weak external field of a given frequency $\omega$.
\begin{eqnarray}
\label{F}
F(t) &=& \eta \left\{ F(\omega) e^{-i\omega t}
                         + F^\dagger(\omega) e^{i\omega t}
                   \right\} , \\
\label{Fomega}
F(\omega)&=& \frac{1}{2}\sum_{\mu\nu} \left\{
   F_{\mu\nu}^{20}(\omega) A_{\mu\nu}^\dagger
 + F_{\mu\nu}^{02}(\omega)  A_{\mu\nu}
 \right\}  \nonumber \\
 && + \sum_{\mu\nu} F_{\mu\nu}^{11}(\omega)  B_{\mu\nu} ,
\end{eqnarray}
where $A_{\mu\nu}^\dagger\equiv a_\mu^\dagger a_\nu^\dagger$
and $B_{\mu\nu}\equiv a_\mu^\dagger a_\nu$.
A small real parameter $\eta$ is introduced for the linearization.
In the small-amplitude limit, the second term ($B$-part)
in Eq. (\ref{Fomega}) can be omitted,
because it doesn't contribute in the linear approximation.
The Bogoliubov transformation of the external fields
($F_{\mu\nu}^{20}(\omega)$ and $F_{\mu\nu}^{02}(\omega)$)
 is given in Appendix~\ref{app-external}.

The external perturbation $F(t)$ induces a density oscillation
around the ground state with the same frequency $\omega$.
The density oscillation, then, produces the induced fields in the
single-particle Hamiltonian, $h(t)=h_0+\delta h(t)$
and in the pair potential, $\Delta(t)=\Delta+\delta\Delta(t)$.
Thus, the Hamiltonian, Eq. (\ref{TDHFB_Hamiltonian}),
is decomposed into the static and oscillating parts;
$H(t)=H_0+\delta H(t)$.
\begin{eqnarray}
\label{delta_H}
\delta H(t) &=& \eta \left\{
\delta H(\omega)e^{-i\omega t}+ \delta H^\dagger(\omega)e^{i \omega t} 
\right\}, \\
\label{delta_H20}
\delta H(\omega)&=& \frac{1}{2}\sum_{\mu\nu} \left\{
\delta H_{\mu\nu}^{20}(\omega) A_{\mu\nu}^\dagger
+\delta H_{\mu\nu}^{02}(\omega) A_{\mu\nu}
\right\} .
\end{eqnarray}
Here, the $B$-part is again neglected in Eq. (\ref{delta_H20}).
See Appendix \ref{app-dH} for the derivation of $\delta H(\omega)$.
Explicit expressions for
$\delta H_{\mu\nu}^{20}(\omega)$ and
$\delta H_{\mu\nu}^{02}(\omega)$ are found in Eqs. (\ref{dH20}) and
(\ref{dH02}), respectively.

The time-dependent quasi-particle operators are decomposed in a
similar manner:
\begin{equation}\label{eq-a}
 a_{\mu}(t) = \{ a_{\mu}+ \delta a_{\mu}(t)\}e^{iE_{\mu}t},  
\end{equation}
where $\delta a_\mu(t)$ can be expanded in the quasi-particle creation
operators:
\begin{equation}\label{deltaa}
 \delta a_{\mu}(t)=\eta \sum_{\nu} a^{\dagger}_{\nu} \left( \frac{}{} X_{\nu \mu}(\omega)e^{-i\omega t} + Y_{\nu \mu}^{*}(\omega) e^{i\omega t}\right) \ .
\end{equation}
It should be noted that $\delta a_\mu$ can be expanded only in terms 
of the creation operators,
because the annihilation operators in the right-hand side of Eq. (\ref{deltaa})
simply represent the transformation among themselves,
$a_\mu(t) = \sum_\nu C_{\mu\nu}(t) a_\nu$, and do not affect
$\rho$ and $\kappa$.
The amplitudes, $X$ and $Y$, must be anti-symmetric to satisfy the fermionic
commutation relation,
$\{ a_\mu(t),a_\nu(t) \} =0$.
Keeping only linear terms in $\eta$, Eq. (\ref{motodist}) becomes
\begin{equation}\label{eqmot}
i \frac{\partial \delta a_{\mu}(t)}{\partial t}  
=E_{\mu}\delta a_{\mu}(t)+[H_{0},\delta a_{\mu}(t)]+[\delta H(t)+F(t), a_{\mu}].
\end{equation}
Substituting Eqs. (\ref{F})-(\ref{deltaa}) into Eq. (\ref{eqmot}),
we obtain the linear-response equations:
\begin{equation}\label{seconda} 
 \left\{ 
\begin{array}{cc}
\DS
\left( E_{\mu}+E_{\nu}- \omega   \right) 
  X_{\mu \nu}(\omega) + \delta H^{20}_{\mu \nu}(\omega)      & = 
 F^{20}_{\mu\nu}(\omega) \\
\DS
\left(  E_{\mu} +E_{\nu}+\omega \right) Y_{\mu \nu}(\omega) 
+ \delta H^{02}_{\mu \nu}(\omega) & = 
 F^{02}_{\mu\nu}(\omega)
\end{array}
\right. .
\end{equation}
In Eq. (\ref{seconda}), setting the frequency complex,
$\omega\rightarrow \omega+i\gamma/2$, we can introduce a smearing with
a width $\gamma$.

Expanding $\delta H^{20}(\omega)$ and $\delta H^{02}(\omega)$
in terms of the forward and backward amplitudes, $X$ and $Y$,
we obtain a familiar expression of the equation \cite{RS}:
\begin{equation}\label{QRPA-standard}
\left[
\!
\left(
 \begin{array}{cc}
  A     &    B     \\
  B^{*} &    A^{*}
 \end{array}
\right)
-\omega
\left(
\begin{array}{cc}
  I\!\!I &      ~~0     \\
   0        &  -I\!\!I    
\end{array}
\right)
\right]
\left(
\begin{array}{c}
 X(\omega)   \\
 Y(\omega)
\end{array}
\right)
\!
=\!
\left(
\!
\begin{array}{c}
 F^{20}(\omega)    \\
 F^{02}(\omega)
\end{array}
\!
\right).
\end{equation}
This matrix formulation requires us to calculate the QRPA matrix
elements, $A_{\mu\nu,\mu'\nu'}$ and $B_{\mu\nu,\mu'\nu'}$.
This is a tedious task and their dimension,
which is equal to the number of two-quasi-particle excitations,
becomes huge especially for deformed nuclei.
Instead, in the FAM \cite{fam}, we keep the form of Eq. (\ref{seconda}) and
calculate the induced fields $\delta H^{20}(\omega)$
and $\delta H^{02}(\omega)$
using the numerical differentiation.
We explain this trick in the next section.

\section{Finite amplitude method for the induced fields} \label{secFAM}

The expressions for $\delta H^{20}$ and $\delta H^{02}$ in Eq. (\ref{seconda})
are given
in Eqs. (\ref{dH20}) and (\ref{dH02}), respectively.
Thus, we need to calculate $\delta h(\omega)$ and
$\delta\Delta^{(\pm)}(\omega)$ for given $X$ and $Y$.
We perform this calculation following the spirit of the FAM \cite{fam}.

From Eqs. (\ref{eq-a}) and (\ref{deltaa}),
we obtain the time-dependent quasi-particle wave functions:
\begin{equation}
\label{phase_factor}
\begin{pmatrix}
U_{\mu}(t) \\
V_{\mu}(t)
\end{pmatrix}
=
\begin{pmatrix}
\mathcal{U}_{\mu}(t) \\
\mathcal{V}_{\mu}(t)
\end{pmatrix}
e^{iE_\mu t}
 ,
\end{equation}
where
\begin{eqnarray}
\label{Ut}
\mathcal{U}_{k\mu}(t) &=& \left\{ U + \eta \left( V^* X^* e^{i\omega t}
                        + V^* Y e^{-i\omega t}\right)\right\}_{k\mu}
 ,\ \ \\
\label{Vt}
\mathcal{V}_{k\mu}(t) &=& \left\{ V + \eta \left( U^* X^* e^{i\omega t}
                        + U^* Y e^{-i\omega t}\right)\right\}_{k\mu}
 .\ \  
\end{eqnarray}

First, let us discuss how to obtain $\delta h(\omega)$.
The time-dependent single-particle Hamiltonian $h(t)$
depends on the densities which
are determined by the wave functions $(U(t),V(t))$.
Therefore, $h(t)$ can be regarded as a functional of wave functions as
\begin{equation}
h\left[U^*(t), V^*(t); U(t),V(t)\right] 
=h\left[\mathcal{U}^*(t), \mathcal{V}^*(t);
        \mathcal{U}(t),\mathcal{V}(t)\right] 
.
\end{equation}
Here, it should be noted that the phase factors,
$e^{iE_\mu t}$ in Eq. (\ref{phase_factor}),
do not play a role.
This is because
$h$ is a functional of densities, $\rho$, $\kappa$, and $\kappa^*$,
which are given by products of one of $(U,V)$ and one of
the complex conjugate $(U^*,V^*)$,
such as $\rho=V^* V^T$ and $\kappa=V^* U^T$.
Therefore, the time-dependent phases in Eq. (\ref{phase_factor})
are always canceled, thus can be omitted.

Now, we take the small-amplitude limit, keeping only the
linear order in $\eta$.
\begin{eqnarray}
h(t)&=&h\left[\mathcal{U}^*(t), \mathcal{V}^*(t);
        \mathcal{U}(t),\mathcal{V}(t)\right]  \nonumber \\
&=& h\left[U^*, V^*; U,V \right] 
+ \eta \left\{ \delta h(\omega)
e^{-i\omega t} + \mbox{h.c.}
\right\} .
\end{eqnarray}
Here, $\delta h(\omega)$ can be obtained using Eqs. (\ref{Ut})
and (\ref{Vt}), expanding up to the first order in $\eta$ and
collecting terms proportional to $e^{-i\omega t}$, as
\begin{eqnarray}
\delta h(\omega) &=&
\frac{\partial h}{\partial U^*} \cdot V X
+ \frac{\partial h}{\partial V^*} \cdot U X \nonumber \\
&& +\frac{\partial h}{\partial U} \cdot V^* Y
+ \frac{\partial h}{\partial V} \cdot U^* Y .
\end{eqnarray}
The calculation of the derivatives,
such as $\partial h_{kl}/\partial U^*_{k'\mu}$,
is a tedious task and requires a large memory capacity for their storage
in the computation.
In the FAM, we avoid this explicit expansion, instead write the
same quantity as follows:
\begin{equation}
\label{dhomega_fam}
\delta h(\omega) =
\frac{h\left[\bar{U}_\eta^*, \bar{V}_\eta^*; U_\eta,V_\eta \right] - h\left[U^*, V^*; U,V \right]}{\eta}
 + \mathcal{O}(\eta^2) ,
\end{equation}
where $\bar{U}_\eta^*$, $\bar{V}_\eta^*$, $U_\eta$, and $V_\eta$ are given by
\begin{equation}
\begin{split}
\bar{U}_\eta^* &\equiv U^* + \eta VX ,\quad \bar{V}_\eta^* \equiv V^* + \eta UX ,
 \\
U_\eta &\equiv U + \eta V^* Y ,\quad V_\eta \equiv V + \eta U^* Y .
\end{split}
\label{UV_eta_p}
\end{equation}
This is the FAM formula for the calculation of $\delta h(\omega)$.
All we need in the computer program is a subroutine to calculate
the single-particle Hamiltonian as a function of the wave functions,
$h[\bar{U}^*,\bar{V}^*;U,V]$.

For the pair potential, basically, the same arguments lead to the
FAM formulae for $\delta\Delta^{(\pm)}$.
The time-dependent pair potential $\Delta(t)$ can be written as
\begin{eqnarray}
\label{Delta_t}
\Delta(t)&=&\Delta\left[\mathcal{U}^*(t), \mathcal{V}^*(t);
        \mathcal{U}(t),\mathcal{V}(t)\right]  \nonumber \\
&=& \Delta\left[U^*, V^*; U,V \right]  \nonumber \\
 && + \eta \left\{ \delta \Delta^{(+)}(\omega)
e^{-i\omega t} + \delta\Delta^{(-)}(\omega) e^{i\omega t}
\right\} .
\end{eqnarray}
Here, $\delta\Delta^{(+)}$ and $\delta\Delta^{(-)}$ are
independent, since $\Delta(t)$ is non-Hermitian in general.
$\delta\Delta^{(+)}$ can be written in the same form
as Eq. (\ref{dhomega_fam}).
\begin{eqnarray}
\label{dDp_fam}
\delta \Delta^{(+)}(\omega) &=&
\frac{\Delta\left[\bar{U}_\eta^*, \bar{V}_\eta^*; U_\eta,V_\eta \right]
 - \Delta\left[U^*, V^*; U,V \right]}{\eta} \nonumber \\
 && + \mathcal{O}(\eta^2) ,
\end{eqnarray}
where $\bar{U}_\eta^*$, $\bar{V}_\eta^*$, $U_\eta$, and $V_\eta$ are given by
Eq. (\ref{UV_eta_p}).

The expression for $\delta\Delta^{(-)}$ is
also obtained from Eq. (\ref{Delta_t}),
collecting terms proportional to $e^{i\omega t}$.
It is given by the same expression as Eq. (\ref{dDp_fam}),
\begin{eqnarray}
\label{dDm_fam}
\delta \Delta^{(-)}(\omega) &=&
\frac{\Delta\left[\bar{U}_\eta^*, \bar{V}_\eta^*; U_\eta,V_\eta \right]
 - \Delta\left[U^*, V^*; U,V \right]}{\eta} \nonumber \\
 && + \mathcal{O}(\eta^2) .
\end{eqnarray}
However, $(\bar{U}_\eta^*,\bar{V}_\eta^*;U_\eta, V_\eta)$ here are
 different from Eq. (\ref{UV_eta_p})
and given by
\begin{equation}
\begin{split}
\bar{U}_\eta^* &\equiv U^* + \eta VY^* ,\quad \bar{V}_\eta^* \equiv V^* + \eta UY^* ,
\\
U_\eta &\equiv U + \eta V^* X^* ,\quad V_\eta \equiv V + \eta U^* X^* .
\end{split}
\label{UV_eta_m}
\end{equation}

The essential trick of the FAM is to calculate the induced fields,
$\delta h(\omega)$ and $\delta\Delta^{(\pm)}$,
according to Eqs. (\ref{dhomega_fam}), (\ref{dDp_fam}), and (\ref{dDm_fam})
with a small but finite parameter $\eta$.
Of course, the $\eta^2$ and higher-order terms bring some numerical errors, but
they are negligible.
Therefore, for given $X$ and $Y$,
we are able to calculate these induced fields,
by using the static HFB code with some minor modifications.
$\delta H^{20}(\omega)$ and $\delta H^{02}(\omega)$ of
Eq. (\ref{seconda}) in the quasi-particle basis can be calculated
with Eqs. (\ref{dH20}) and (\ref{dH02}), respectively.
Then, we may solve the QRPA linear-response equation (\ref{seconda})
to obtain the self-consistent amplitudes, $X$ and $Y$,
utilizing an iterative algorithm
(See Sec. \ref{sec:fammethod}).

\subsection{Induced fields in terms of densities}

Although the basic formulae of the FAM has been provided in Sec. \ref{secFAM},
we may need to modify them in the practical implementation of the FAM. 
For instance, some HFB codes, such as HFBRAD,
contain subroutines to calculate mean fields
as functions of densities, not of wave functions.
In this subsection, we rewrite Eqs. (\ref{dhomega_fam}),
(\ref{dDp_fam}), and (\ref{dDm_fam}) in terms of densities.

The density $\delta\rho(t)$ is written up to linear order in $\eta$ as
\begin{equation}\label{eq-t-dep-rho}
 \begin{split}
 \rho(t) & =V^{*}(t)V^{T}(t)  \\
         & =\rho_0 + \eta \left( \delta\rho(\omega)e^{-i\omega t}
           + \mbox{h.c.}  \right) ,
 \end{split}
\end{equation}
where
\begin{equation}
 \delta \rho(\omega) = UXV^{T}+V^{*}Y^{T}U^{\dagger} . \label{ddensfreq}
 \end{equation}
This can be written in the FAM form:
\begin{eqnarray}
 \delta \rho(\omega) &=& \frac{\rho_\eta - \rho_0}{\eta}
 + \mathcal{O}(\eta^2) \nonumber \\
\label{eq-rho-eta} 
 &=& \frac{\bar{V}_\eta^* V_\eta^T - V^*V^T}{\eta}
 + \mathcal{O}(\eta^2) ,
 \end{eqnarray}
where $\bar{V}_\eta^*$ and $V_\eta$ are given in Eq. (\ref{UV_eta_p}).

The pair tensor $\kappa(t)$, which is non-Hermitian, can be expressed in
a similar manner.
\begin{equation}\label{eq-t-dep-kappa}
 \begin{split}
 \kappa(t) & =V^{*}(t)U^{T}(t) \\
& =\kappa_0  +\eta \left( \delta\kappa^{(+)} e^{-i\omega t}
       + \delta\kappa^{(-)} e^{i \omega t} \right) .
\end{split}
\end{equation}
Here, $\kappa^{(\pm)}$ can be given in the explicit form as
\begin{eqnarray}
 \delta \kappa^{(+)}(\omega)           & = UXU^{T}+V^{*}Y^{T}V^{\dagger}, \label{dkfreq}\\
 \delta \kappa^{(-)}(\omega) & = V^{*}X^{\dagger}V^{\dagger}+ UY^{*}U^{T}, \label{dkfreq2}
\end{eqnarray}
and in the FAM form as
\begin{eqnarray}
 \delta \kappa^{(\pm)}(\omega) &=& \frac{\kappa_\eta^{(\pm)} - \kappa_0}{\eta}
 + \mathcal{O}(\eta^2)  \nonumber \\
 &=& \frac{\bar{V}_\eta^* U_\eta^T - V^*U^T}{\eta}
 + \mathcal{O}(\eta^2) ,
\label{eq-kappa-eta}
 \end{eqnarray}
where $\bar{V}_\eta^*$ and $U_\eta$ are given in Eq. (\ref{UV_eta_p}) for
$\kappa_\eta^{(+)}$ while they are given by Eq. (\ref{UV_eta_m}) for
$\kappa_\eta^{(-)}$.

Now, let us present how to obtain the induced fields in terms of the
densities.
In general, $h(t)$ and $\Delta(t)$ may depend on
$\rho$, $\kappa$, and $\kappa^*$.
\begin{equation}
h(t)=h\left[\rho(t),\kappa(t),\kappa^*(t)\right], \quad
\Delta(t)=\Delta\left[\rho(t),\kappa(t),\kappa^*(t)\right] .
\end{equation}
In order to obtain the induced fields, all we need to do is to replace
$\rho$ by $\rho_\eta$ defined in Eqs. (\ref{eq-rho-eta}),
and $\kappa$ by $\kappa_\eta^{(\pm)}$ in Eq. (\ref{eq-kappa-eta}), as follows:
\begin{eqnarray}
\delta h(\omega) &=&
\frac{h\left[\rho_\eta,\kappa_\eta^{(+)},\kappa_\eta^{(-)*} \right]
- h\left[\rho,\kappa,\kappa^* \right]}{\eta}
 , \\
\delta \Delta^{(+)}(\omega) &=&
\frac{\Delta\left[\rho_\eta,\kappa_\eta^{(+)},\kappa_\eta^{(-)*} \right]
- \Delta\left[\rho,\kappa,\kappa^* \right]}{\eta}
 , \\
\delta \Delta^{(-)}(\omega) &=&
\frac{\Delta\left[\rho_\eta,\kappa_\eta^{(-)},\kappa_\eta^{(+)*} \right]
- \Delta\left[\rho,\kappa,\kappa^* \right]}{\eta}
 ,
\end{eqnarray}
where the terms of the second and higher orders in $\eta$ are neglected.

\section{Summary of The Finite Amplitude Method}\label{sec:fammethod}
Here we provide a summary of the FAM for the QRPA linear-response
calculation for a prompt application.
Later, we discuss applications of the FAM to the Skyrme functionals,
however, the FAM formulated in this and previous sections
is applicable to any kind of energy density functional
(mean-field) models.

\subsection{Numerical procedure}

The aim is to solve the linear-response equation (\ref{seconda})
for a given external field $F$.
In order to obtain the forward and backward amplitudes, $X$ and $Y$,
we resort to an iterative algorithm.
Namely, we start from the initial guess for
$(X,Y)=(X^{(0)},Y^{(0)})\equiv\vec{x}^{(0)}$,
and calculate $\delta h(\omega)$ and $\delta\Delta^{(\pm)}(\omega)$
according to the formulae,
(\ref{dhomega_fam}), (\ref{dDp_fam}), and (\ref{dDm_fam}).
Then, they are converted into
$\delta H^{20}(\omega)$ and $\delta H^{02}(\omega)$,
using Eqs. (\ref{dH20}) and (\ref{dH02}), respectively.
In this way, we can evaluate the left and right hand sides of
Eq. (\ref{seconda}) for a given $(X,Y)$.

Since Eq.~(\ref{seconda}) is equivalent to Eq.~(\ref{QRPA-standard}),
it is a linear algebraic equation for the vector
$\vec{x}\equiv (X,Y)$, in the form of $A\vec{x}=\vec{b}$. 
Many different algorithms are available for the solution of linear systems.
In this paper, we resort to a procedure based on Krylov spaces called
generalized conjugate residual (GCR) method \cite{Saad}.
Within these kinds of methods, a succession of approximate solutions
($\vec{x}^{(0)}, \vec{x}^{(1)}, \vec{x}^{(2)}, \cdots$)
converging to the exact one is obtained by the iteration.
The GCR algorithm consists in a series of steps each containing
the operation of the matrix $A$ on a given vector, and
sums and scalar products of two vectors.
For the given $\vec{x}=(X,Y)$, $A\vec{x}$ is equal to the left hand side of
Eq. (\ref{seconda}).
Therefore, the quantity $A\vec{x}$ can be calculated without the explicit
knowledge of the QRPA matrix itself.

Here, we summarize the formulae.
The linear response equation is given by $A\vec{x}=\vec{b}$, where
$$
\vec{x}\equiv
\begin{pmatrix}
X_{\mu\nu} \\
Y_{\mu\nu}
\end{pmatrix}
, \quad
\vec{b}\equiv
\begin{pmatrix}
F_{\mu\nu}^{20} \\
F_{\mu\nu}^{02}
\end{pmatrix} ,
$$
and 
\begin{equation*}
A\vec{x}= \begin{pmatrix}
\left( E_{\mu}+E_{\nu}- \omega   \right) 
  X_{\mu \nu}(\omega) + \delta H^{20}_{\mu \nu}(\omega) \\
\left(  E_{\mu} +E_{\nu}+\omega \right) Y_{\mu \nu}(\omega) 
+ \delta H^{02}_{\mu \nu}(\omega)
\end{pmatrix} ,
\end{equation*}
where
\begin{eqnarray*}
\delta H^{20}_{\mu\nu}(\omega) &=&
  U^{\dagger} \delta h             V^{*}
- V^{\dagger} \delta \Delta^{(-)*} V^{*}  \nonumber \\
&+& U^{\dagger} \delta \Delta^{(+)}  U^{*}
- V^{\dagger} \delta h^T           U^{*} , \\
\delta H^{02}_{\mu\nu}(\omega) &=&
 - V^T \delta h             U
 + U^T \delta \Delta^{(-)*} U   \nonumber \\
&-& V^T \delta \Delta^{(+)}  V
 + U^T \delta h^T           V  .
\end{eqnarray*}
Denoting $h$ and $\Delta$ collectively as
$\mathcal{H}\equiv(h, \Delta)$,
the induced fields $\delta\mathcal{H}$ are calculated by the FAM formulae,
\begin{equation}
\label{induced_fields}
\delta \mathcal{H} =
\frac{\mathcal{H}\left[\bar{U}_\eta^*, \bar{V}_\eta^*; U_\eta,V_\eta \right]
 -\mathcal{H}\left[U^*, V^*; U,V \right]}{\eta} , \\
\end{equation}
where $(\bar{U}_\eta^*, \bar{V}_\eta^*; U_\eta, V_\eta)$ are given by
\begin{eqnarray*}
\bar{U}_\eta^* &\equiv& U^* + \eta VY^* ,\quad \bar{V}_\eta^* \equiv V^* + \eta UY^* , \\
U_\eta &\equiv& U + \eta V^* X^* ,\quad V_\eta \equiv V + \eta U^* X^* .
\end{eqnarray*}
for the calculation of $\delta h(\omega)$ and $\delta\Delta^{(+)}$.
For $\delta\Delta^{(-)}$,
they are
\begin{eqnarray*}
\bar{U}_\eta^* &\equiv& U^* + \eta VX ,\quad \bar{V}_\eta^* \equiv V^* + \eta UX ,
\nonumber \\
U_\eta &\equiv& U + \eta V^* Y ,\quad V_\eta \equiv V + \eta U^* Y .
\end{eqnarray*}
The final result does not depend on the parameter $\eta$, as far as
it is in a reasonable range.
The choice of $\eta$ is discussed in Sec. \ref{sectComparison}.

\subsection{Calculation of the strength function}

Using the solution $(X,Y)$, we can calculate the
strength function following the same procedure as Ref. \cite{fam}.
\begin{equation}
 \frac{dB (\omega; F) }{d\omega}
\equiv \sum_{n>0} |\langle n| F | 0 \rangle|^2 \delta (\omega-E_n)
= -\frac{1}{\pi} \mbox{Im} S(F;\omega).
\end{equation}
Here, $S(F;\omega)$ is obtained from the solution $(X,Y)$.
For the operator in the form of Eq. (\ref{F-type}),
we may calculate $S(F;\omega)$ as
\begin{equation}
\label{SF-F-type}
 S(F;\omega) =  \mbox{tr}\left\{ f^{\dagger} \delta \rho(\omega) \right\},
\end{equation}
For the operator in the form of Eq. (\ref{G-type}),
we have
\begin{equation}
\label{SF-G-type}
 S(F;\omega) =  \mbox{tr}\left\{ g^\dagger \delta \kappa^{(+)}(\omega)
                    + g^{'\dagger} \delta \kappa^{(-)*}(\omega) \right\} .
\end{equation}
For both cases, in the two-quasi-particle basis,
Eqs. (\ref{SF-F-type}) and (\ref{SF-G-type}) can be written in
the unified expression.
\begin{equation}
 S(F,\omega)
=  \frac{1}{2} \sum_{\mu\nu} \left\{ F_{\mu\nu}^{20*} X_{\mu\nu}(\omega) 
                          + F_{\mu\nu}^{02*} Y_{\mu\nu}(\omega) \right\} ,
\end{equation}
where $F^{20}$ and $F^{02}$ are given by Eqs. (\ref{F20-F-type})
and (\ref{F02-F-type}) for the former case,
and by Eqs. (\ref{F20-G-type})
and (\ref{F02-G-type}) for the latter.

\section{Application of the FAM to the HFBRAD}\label{sectComparison}
In order to assess the validity of the FAM,
we install the FAM in the HFBRAD code \cite{BennaDoba}.
It has to be noted that the formalism of the HFBRAD is slightly different
from the one used in this paper which follows the notations in Ref. \cite{RS}.
In particular, the wave functions $(\varphi_{1\mu},\varphi_{2\mu})$,
the pairing tensor $\tilde\rho$, and the pair potential $\tilde h$
are defined in a different manner;
$\varphi_{1\mu}(k)=U_{k\mu}$,
$\varphi_{2\mu}(k)=V_{\bar{k}\mu}$,
$\tilde\rho_{kl}=\kappa_{k\bar{l}}$, and
$\tilde{h}_{kl}=\Delta_{k\bar{l}}$,
where $\bar{k}$ is the time-reversal state of $k$.
A detailed discussion on the difference among the two notations
can be found in Ref. \cite{Doba1}.

The HFBRAD \cite{BennaDoba} is a well known code which solves the HFB
in the radial coordinate space assuming the spherical symmetry.
It has been designed to provide fast and reliable solutions for the
ground state of spherical even-even nuclei.
For these nuclei, the time-odd densities are identically zero
and thus they have not been implemented in the code.
In order to render the QRPA fully self-consistent, we have to add the
time-odd terms in the calculation of the induced fields.
This task can be simplified for a case of the presence of spherical 
and space-inversion symmetry, such as in the case of monopole excitations.
For this case, the only time-odd terms with non-zero contribution are
those due to the current density \cite{DobaSimOdd},
moreover the only non-vanishing component of the current density is radial.

We calculate the strength function of the isoscalar monopole for 
a neutron-rich nucleus, $^{174}$Sn.
To check the self-consistency by looking at the spurious component,
we also calculate the strength of the
nucleon number operator.
Both operators are given by the form of Eq. (\ref{F-type})
with
$f_{kl}=\langle k|r^{2}|l\rangle$ for the isoscalar monopole operator
and $f_{kl}=\delta_{kl}$ for the number operator.
    
In order to obtain the strength function, first,
we have to solve the HFB equations to construct the
ground-state wave functions $(U,V)$.
It is accomplished by using the HFBRAD code.
The parameters of the present calculation are adjusted to the values
used by Terasaki and co-workers in Ref. \cite{Terasaki1};
The box size is $R_{box}=20$ fm, the quasi-particle energy cutoff is
$E_{\rm qp}^{\rm c}=200$ MeV,
the maximum angular momenta of the quasi-particle states are
$j_{\rm max}^{n}=21/2$ for neutrons, and $j_{\rm max}^{p}=15/2$ for protons.
We use the Skyrme functional with the SkM* parameter set \cite{SkM*}
in the ph-channel and a delta interaction of the volume type
with the strength $V_0 = -77.5$ MeV fm$^{3}$ for the pp- and hh-channels.

The next step is solving the linear-response equation for a given
external field of the frequency $\omega$.
At first, we build the induced fields, $\delta h(\omega)$ 
and $\delta\Delta^{(\pm)}(\omega)$,
starting from a guess choice of the QRPA amplitudes $(X^{(0)},Y^{(0)})$,
according to Eq. (\ref{induced_fields}).
In the present calculation, we choose either $X^{(0)}=Y^{(0)}=0$ or
the values of $X$ and $Y$ at the previous energy $\omega$ calculated.
We resort to the iterative algorithm of the GCR method to solve the equation
(\ref{seconda}).
We include all the two-quasi-particle states $(\mu\nu)$ within the
HFB model space defined above ($E_{\mu(\nu)} \leq 200$ MeV).
The two-quasi-particle space amounts to 12,632 states for $J^\pi=0^+$.
Note that this number becomes much larger if we treat deformed systems.
We set the accuracy of the convergence to be $\epsilon<10^{-5}$, where
$\epsilon\equiv \|A\vec{x}-\vec{b}\|^{2}/\|\vec{b}\|^{2}$.
The number of iterations needed
depends on $\omega$; at low energies, about 50-60 iterations are enough
to reach the convergence, while, close to the central peak at 12 MeV,
more than 300 iterations are needed.

  \begin{table}[t]
     \begin{center}
        \begin{tabular}{|c|c|c|c|c|c|c|}
        \hline   
\multicolumn{7}{|c|}{$^{174}$ Sn, $0^{+}$  }\\
        \hline 
\multicolumn{1}{|c|}{              } &
\multicolumn{2}{|c|}{ $\omega=4$ MeV  } &
\multicolumn{2}{|c|}{ $\omega=12$ MeV  } &
\multicolumn{2}{|c|}{ $\omega=20$ MeV  }   \\  
        \hline
         $\eta$  &     $\epsilon$   &  $N_{\rm iter}$ &   $\epsilon$  & $N_{\rm iter}$     &    $\epsilon$&  $N_{\rm iter}$\\
        \hline
      $10^{-2}$  & 0.44                & 1000  &  1.63 $\cdot 10^{-1}$  &1000  &  $8.84 \cdot 10^{-3}$ & 1000  \\  
        \hline                                                                                               
       $10^{-4}$ &  $6.10\cdot 10^{-5}$& 1000  &  1.76 $\cdot 10^{-5}$  &1000  &   $< 10^{-5}$         &  469  \\  
        \hline                                                                                               
       $10^{-5}$ &  $< 10^{-5}$        &  161  &       $< 10^{-5}$      & 439  &   $< 10^{-5}$         &  469  \\  
        \hline                                                                                               
      $10^{-8}$  &$<  10^{-5}$         &  161  &       $<  10^{-5}$     & 439  &  $< 10^{-5}$          &  469  \\
        \hline                                                                                               
     $10^{-9}$   & $<  10^{-5}$        &  161  &       $< 10^{-5}$      & 439  &   $< 10^{-5}$         &  469  \\
        \hline                                                                                               
     $10^{-10}$  &$<  10^{-5}$         &  161  &  1.19  $\cdot 10^{-5}$ &1000  &  $1.46 \cdot 10^{-5}$ & 1000  \\
        \hline 
        \end{tabular}
     \end{center} 
\caption{Convergence properties of the calculation.
The obtained accuracy
$\epsilon=\|A\vec{x}-\vec{b}\|^{2}/\|\vec{b}\|^{2}$ and
the number of GCR iteration $N_{\rm iter}$ to reach $\epsilon<10^{-5}$
are shown for different values of $\eta$.
The initial vector is chosen as $\vec{x}^{(0)}=(X^{(0)},Y^{(0)})=(0,0)$ 
and
the maximum number of iterations is set at $N_{\rm iter}=1,000$.
} 
\label{convergence2}
\end{table}

We studied the convergence quality of the solutions as a function of the parameter $\eta$ used for the numerical derivative.
This is shown in Table \ref{convergence2}.
If $\eta$ is too big ($\eta \geq 10^{-4}$) the derivative of the FAM becomes inaccurate and the linearity of the procedure is partially broken.
The residue $\|A\vec{x}-\vec{b}\|$ reaches a plateau where increasing the number of iterations cannot improve it anymore.
For $10^{-5}\leq \eta \leq 10^{-9}$, the calculations converge well
and the resulting strength function is stable.
If $\eta$ becomes smaller than $~10^{-10}$, the numerical precision limits are reached and the GCR procedure can no longer obtain the required precision. 
Therefore, we may conclude that the parameter $\eta$ in the range of
$10^{-5}\leq \eta \leq 10^{-9}$ is appropriate to obtain the induced fields
accurately.
Although the constant value $\eta=10^{-8}$ is adopted in this paper,
we may use a more sophisticated choice,
such as the $\omega$-dependent $\eta$ values \cite{fam,Inakura},

We report the strength function of the isoscalar monopole mode.
To smear the strengths at discrete eigenenergies,
we add an imaginary term to the energy:
$\omega \rightarrow \omega + i\gamma/2$, where $\gamma = 1.0$ MeV.
This procedure is almost equivalent to smearing the strength function
with a Lorentzian function with a width equal to $\gamma$. 
The calculated energy-weighted strengths are summed up to $300$ MeV
and we found that they exhaust 99.6 \% of the theoretical sum-rule value
given by
$\frac{ 2 }{m} A \langle r^{2}\rangle$.
\begin{figure}
\centering
\includegraphics[width=8.5cm]{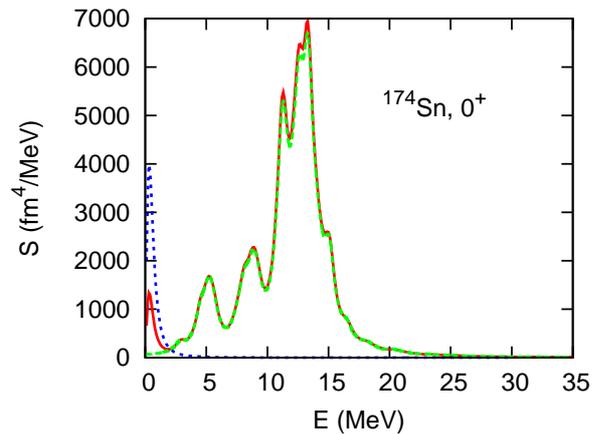}
\caption{(Color online)
Calculated transition strength of the isoscalar monopole $0^{+}$
excitations in $^{174}$Sn (solid red curve), compared with the result
in \cite{Terasaki1} with the cutoff (iii) (green dashed curve).
The transition strength associated to the number operator, magnified by
a factor of 10,000, in units of MeV$^{-1}$ is shown by the blue dotted
curve.
See text for details.}
 \label{fig:noi}
\end{figure}

In Fig. \ref{fig:noi}, we compare our results (solid red curve) with
the one in Ref. \cite{Terasaki1} (dashed green curve).
The self-consistent result obtained by Terasaki et al.  \cite{Terasaki1}
also employs the HFB solutions calculated with the HFBRAD.
However, in Ref. \cite{Terasaki1},
the QRPA matrix is calculated in the canonical-basis representation and
an additional truncation
of the two-quasi-particle space is introduced for
the construction of the QRPA matrix .
In contrast, we introduce no additional truncation for our FAM calculation.
We compare our results with the one of the cutoff (iii) in Ref. \cite{Terasaki1}
which takes into account the highest number of states for the construction
of the QRPA matrix;
all the proton quasi-particles up to 200 MeV
and the neutron canonical levels with occupancy $v^{2}> 10^{-16}$.

In the first two peaks at $E\sim 5$ and $8.5$ MeV,
the two curves are almost perfectly overlapping. 
The peaks between 11 MeV and 18 MeV occur at the same energy
for the two calculations while their height is slightly different. 
The bump close to zero energy resulting in our calculations has
to be attributed to the presence of a spurious mode.
To check the position of the spurious mode related to the pairing rotation
of the neutrons,
we included in Fig. \ref{fig:noi} the transition strength associated
to the number operator, by the blue dashed line.
The spurious mode is well localized close to zero energy.

The present result demonstrates the accuracy and usefulness of the FAM
for the superfluid systems.
Even if the two codes include
some differences in the truncation of the two-quasi-particle space,
the similarity of the results is very satisfying

\section{Conclusions}\label{conclusioni}
The finite amplitude method for the QRPA has been presented.
The basic idea is identical to the original FAM \cite{fam},
that we resort to a numerical differentiation to calculate
the induced fields and then solve the linear-response equation
with an iterative algorithm such as the GCR.
With the FAM, a HFB code with simple modifications can be turned
into a QRPA code.
Especially, it is very easy to construct the QRPA code which
has the same symmetry of the parent HFB one whose subroutines
are used to perform the numerical derivative.
All the terms present in the TDHFB calculation, including the
time-odd mean fields,
should be taken into account to construct fully self-consistent codes.
This requires us some effort to update the original HFB code.
Still, the necessary task for coding the FAM is much less than that for
the explicit calculation of the QRPA matrix elements
for realistic energy functionals.
In addition, it does not require a large memory capacity, since we do
not construct the QRPA matrix.
We have built a fully self-consistent QRPA code using the HFBRAD
 \cite{BennaDoba}.
The iterative algorithm, for which we adopted the GCR method in this
paper, may be replaced by a better algorithm in future.
The resulting strength functions of the isoscalar $0^{+}$ mode
of $^{174}$Sn show high similarity with the fully self-consistent
calculations in Ref. \cite{Terasaki1}.
Thus, this paper showed the first application of the FAM for
superfluid systems and demonstrated the usefulness of
the FAM for the construction of the QRPA code
by modifying existing HFB codes.

\section*{Acknowledgments}
This work is supported by Grant-in-Aid for Scientific Research(B)
(No. 21340073) and on Innovative Areas (No. 20105003).
We thank the JSPS Core-to-Core Program ``International
Research Network for Exotic Femto Systems''.
We are thankful to J. Terasaki for providing the numerical results of
Ref. \cite{Terasaki1}. 
P.A. thank K. Matsuyanagi for the fruitful discussion on the linear expansion, C. Losa and A.~Pastore for the suggestions on the HFBRAD code and K. Yoshida and T. Inakura for the discussions on the QRPA
and J. Dobaczewski and V. Nesterenko for the useful suggestions.
T.N. thank M. Matsuo for useful discussion and
the support from the UNEDF SciDAC
collaboration under DOE grant DE-FC02-07ER41457.
The numerical calculations were performed in part
on RIKEN Integrated Cluster of Clusters (RICC).

\appendix
\section{Bogoliubov transformation of one-body fields}\label{app:HFBham}

\subsection{Induced fields $\delta H$}\label{app-dH}
 
The TDHFB Hamiltonian is given by Eq. (\ref{TDHFB_Hamiltonian}).
We consider the small-amplitude limit,
$H(t)=H_{0}+\delta H(t)$,
where $H_0$ is the HFB Hamiltonian of Eq. (\ref{HFB_Hamiltonian})
and
\begin{equation}
\label{deltaHdit}
\delta H(t) = \frac{1}{2}
\begin{pmatrix}
c^{\dagger} & c
\end{pmatrix}
\begin{pmatrix}
 \delta h(t)     &    \delta\Delta(t) \\
 -\delta \Delta^*(t) & -\delta h^{*}(t)
\end{pmatrix}
\begin{pmatrix}
 c \\
 c^{\dagger}
\end{pmatrix}
.
\end{equation}
Here, $\delta h(t)$ and $\delta\Delta(t)$ are oscillating as
\begin{eqnarray}
 \delta h(t) & = & \eta \left( 
 \delta h(\omega)e^{-i \omega t} + \delta h^\dagger(\omega) e^{i \omega t} \right) \label{dht} ,   \\
 \delta \Delta(t) & = & \eta \left(
 \delta \Delta^{(+)}(\omega)e^{-i \omega t} + \delta \Delta^{(-)}(\omega)e^{i \omega t} 
                              \right) \label{dDt} .                                                    
\end{eqnarray}
Note that $\delta\Delta^{(\pm)}(\omega)$ are anti-symmetric but
$\delta h(\omega)$ is not necessarily Hermitian.
The induced Hamiltonian, Eq. (\ref{deltaHdit}), is now
expressed in the form of Eq. (\ref{delta_H})
with $\delta H(\omega)$ given by
\begin{equation}
\label{deltaHomega}
\delta H(\omega) = \frac{1}{2}
\begin{pmatrix}
c^{\dagger} & c
\end{pmatrix}
\begin{pmatrix}
 \delta h     &    \delta\Delta^{(+)} \\
 -\delta \Delta^{(-)*} & -\delta h^{T}
\end{pmatrix}
\begin{pmatrix}
 c \\
 c^{\dagger}
\end{pmatrix}
.
\end{equation}
Hereafter, $\delta h(\omega)$ and $\delta\Delta^{(\pm)}(\omega)$ 
are denoted by 
$\delta h$ and $\delta\Delta^{(\pm)}$, for simplicity.

Since the Bogoliubov transformation can be written
in terms of the unitary matrix $\mathcal{W}$ \cite{RS} as follows:
\begin{equation}
\label{W}
\begin{pmatrix}
a \\ a^\dagger
\end{pmatrix}
=
\begin{pmatrix}
U^\dagger & V^\dagger \\
V^T & U^T 
\end{pmatrix}
\begin{pmatrix}
c \\ c^\dagger
\end{pmatrix}
\equiv \mathcal{W^\dagger}   
\begin{pmatrix}
c \\ c^\dagger
\end{pmatrix} ,
\end{equation}
we may rewrite Eq. (\ref{deltaHomega}) in the quasi-particle basis:
\begin{equation}
\label{deltaHomega_2}
\delta H(\omega) = \frac{1}{2}
\begin{pmatrix}
a^{\dagger} & a
\end{pmatrix}
\mathcal{W}^\dagger
\begin{pmatrix}
 \delta h     &    \delta\Delta^{(+)} \\
 -\delta \Delta^{(-)*} & -\delta h^{T}
\end{pmatrix}
\mathcal{W}
\begin{pmatrix}
 a \\
 a^{\dagger}
\end{pmatrix}
.
\end{equation}
This transformation should provide $\delta H^{20}$ and $\delta H^{02}$
in Eq. (\ref{delta_H20}).
\begin{equation}
\label{deltaHomega_3}
\begin{pmatrix}
 \delta H^{11}  & \delta H^{20} \\
 -\delta H^{02} & -(\delta H^{11})^T
\end{pmatrix}
=
\mathcal{W}^\dagger
\begin{pmatrix}
 \delta h     &    \delta\Delta^{(+)} \\
 -\delta \Delta^{(-)*} & -\delta h^{T}
\end{pmatrix}
\mathcal{W}
.
\end{equation}
We write here their explicit expression:
\begin{eqnarray}    \label{dH20}
\delta H^{20}_{\mu\nu}(\omega) &=&
  \left(
  U^{\dagger} \delta h             V^{*}
- V^{\dagger} \delta \Delta^{(-)*} V^{*} \right. \nonumber \\
&&\quad  \left.+ U^{\dagger} \delta \Delta^{(+)}  U^{*}
- V^{\dagger} \delta h^T           U^{*} \right)_{\mu\nu} ,
\end{eqnarray}
\begin{eqnarray}\label{dH02}
\delta H^{02}_{\mu\nu}(\omega) &=&
 \left( - V^T \delta h             U
 + U^T \delta \Delta^{(-)*} U  \right. \nonumber \\
&&\quad \left. - V^T \delta \Delta^{(+)}  V
 + U^T \delta h^T           V \right)_{\mu\nu}.
\end{eqnarray}

\subsection{External field $F$}\label{app-external}
The one-body field in general can be written in a form of Eq. (\ref{Fomega})
in terms of the quasi-particle operators, neglecting a constant.
Suppose that $F(\omega)$ in Eq. (\ref{F}) has a form
\begin{equation}
\label{F-type}
F=\sum_{kl} f_{kl} c_k^\dagger c_l
= \frac{1}{2}
\begin{pmatrix}
c^{\dagger} & c
\end{pmatrix}
\begin{pmatrix}
 f     &    0 \\
 0 & -f^{T}
\end{pmatrix}
\begin{pmatrix}
 c \\
 c^{\dagger}
\end{pmatrix}
,
\end{equation}
where the difference of a constant shift is neglected.
Here, the matrix $f_{kl}$ is a general complex matrix,
since $F(\omega)$ is non-Hermitian in general.
The Bogoliubov transformation as in Eq. (\ref{deltaHomega_3}),
then, leads to $F^{20}$ and $F^{02}$ in Eq. (\ref{Fomega}),
\begin{eqnarray}
\label{F20-F-type}
 F^{20}_{\mu\nu} &=& \left(
                         U^{\dagger} f V^{*} 
                        -V^{\dagger} f^T U^{*}
                   \right)_{\mu\nu} 
,               \\
\label{F02-F-type}
 F^{02}_{\mu\nu} &=& \left(
                        U^T f^T V
                        -V^T f U
                   \right)_{\mu\nu}
.
\end{eqnarray}
In case that $F(\omega)$ has a form of pairing-type
\begin{equation}
\label{G-type}
F= \frac{1}{2} \sum_{kl} \left(
            g_{kl} c_k^\dagger c_l^\dagger
          + g'_{kl} c_l c_k \right) ,
\end{equation}
the same calculation provides $F^{20}$ and $F^{02}$ by
\begin{eqnarray}
\label{F20-G-type}
 F^{20}_{\mu\nu} &=& \left(
                         U^{\dagger} g U^{*} 
                        -V^{\dagger} g' V^{*}
                   \right)_{\mu\nu} 
                   ,  \\
\label{F02-G-type}
 F^{02}_{\mu\nu} &=& \left(
                         U^T g' U
                        -V^T g V
                   \right)_{\mu\nu} .
\end{eqnarray}
%

%

\end{document}